\documentclass[twocolumn,showpacs,amsmath,amssymb,longbibliography,superscriptaddress]{revtex4-1}

\usepackage{color,longtable}
\usepackage{graphicx}
\usepackage{amssymb,amsfonts,amsmath}
\usepackage{here}

\begin{document}
\title{Initial perturbations dependence of non-equilibrium continuous and discontinuous pattern transition}
\author{Rikuya Ishikawa and Rei Kurita}

\affiliation{
Department of Physics, Tokyo Metropolitan University, 1-1 Minamioosawa, Hachiouji-shi, Tokyo 192-0397, Japan
}
\date{\today}
\begin{abstract}
A phase separation in a spatially heterogeneous environment is closely related to intracellular science and material science. 
For the phase separation, initial heterogeneous perturbations play an important role in pattern formations.
In this study, a pattern transition from a lamellar pattern to a columnar pattern is investigated in the presence of a slit pattern as the initial perturbations.
Here it is found that the transition behavior depends on the initial slit width. 
When the initial slit width is close to the width of the columnar pattern at the steady state, the pattern transition is the second-order-like (continuous) transition.  
Meanwhile, the pattern transition becomes the first-order-like (discontinuous) transition if the width of the initial slit is much larger than that at the steady state. 
Then those transition behaviors can be explained by the dynamical path during the pattern formation. 
This finding will advance understanding of the initial perturbation dependence of nonequilibrium phenomena.
\end{abstract}


\maketitle

\section{Introduction}
A phase separation (PS) in which two phases are separated from a mixture is important in a wide range of fields such as material performance and biological activities~\cite{Cahn1958, Hyman2014, Cross1993, Onuki2002, Hamley}.
For example, it is known that SnPb alloy used as solder exhibits phase separation and the formation of phase-separated domains may facilitate the formation of cracks at phase boundaries, resulting mechanical failure~\cite{Dreyer2001}.
Moreover, it has been reported that protein droplets formed by the liquid-liquid phase separation (LLPS) in cells play important roles in transcription~\cite{Hnisz2017}, signal transduction~\cite{Su2016}, and the pathogenesis of neurodegenerative diseases~\cite{Wegmann2018, Alberti2019}. 

Phase separation in binary mixtures has long been studied as a classical problem, and the dynamics of phase separation appears to be well understood both experimentally and theoretically~\cite{Cahn1958, Onuki2002}. 
However, the universality of the dynamics holds for spatially uniform systems, but not for systems with inhomogeneous temperatures or concentrations.
For example, in the case of an inhomogeneous system with large initial concentration fluctuations, a concentric pattern is formed around high concentration areas~\cite{Furukawa1994}. 
In the case of non-stationary temperature fields such as directional quenching (DQ)~\cite{Furukawa1992, Krekhov2009, Tsukada2020, Ishikawa2022}, a random droplet, lamellar and columnar pattern are formed depending on a migration speed $V$ of a quenching front. 
There are many other examples for showing specific pattern formations in systems such as PS with radial quenching~\cite{Kurita2017, Tsukada2019, Tsukada2019-2}, PS with temperature gradient~\cite{Sutapa2018, Sutapa2018-2}, PS with lamination~\cite{Ishikawa2023}, PS with double quenching~\cite{Tanaka1998}, PS with containing particles~\cite{Tanaka1994, Araki2015, Araki2006, Krekhov2013}, particles creation~\cite{Adachi2021}, self-propelled systems~\cite{Cates2015}, and nonreciprocal interaction systems~\cite{Saha2020}. 
Understanding of the phase separation phenomena under such inhomogeneous conditions is an urgent issue since the pattern formation are related with development of new functional materials and LLPS in cells occurs in inhomogeneous concentration and temperature.

Although the importance of initial perturbations has been suggested for the pattern formation, there are few quantitative studies on the effects of the initial perturbations on a transition point and a transition behavior. 
Understanding the effects of the initial perturbations that cause transitions will not only facilitate the understanding of nonequilibrium systems, but also control the transitions which will be useful for applications in various fields. 
Therefore, we investigated the effect of the initial perturbations for the transition from columnar pattern to lamellar pattern in DQ. 
In this study, we prepared a slit pattern into the lamellar pattern as initial perturbations and studied a process of the transition from the columnar pattern to the lamellar pattern. 
It was found that the transition behavior depends on the slits width $h$ and it means that the condition of the initial perturbation plays a critical role for the pattern formation. 

\section{Methods}
In this study, we used modified-Cahn-Hilliard equation for understanding a dynamics of two dimensional phase separation under inhomogeneous temperature fields~\cite{Cahn1958, Jaiswal2013}. 
The normalized equation of the modified-Cahn-Hilliard equation is given as 
\begin{eqnarray}
\frac{\partial \phi}{\partial t} =\nabla ^2 [ \epsilon(x, t)\phi +\phi^3 - \nabla^2 \phi ] \label{CH}
\end{eqnarray} 
where $\phi, t$ and $\epsilon$ are the normalized concentration, the time normalized by the diffusion time and the temperature normalized by the quench depth, respectively. 
The length is normalized by the correlation length. 
We note that the effect of temperature gradient (Ludwig-Soret effect) was neglected. 
When $\epsilon \ge 0$, the mixed state is stable; when $\epsilon < 0$, the mixture is separated into two phases. 
In this study, we define $\phi = 1$ and $\phi = -1$ which are the concentrations after phase separation as A (white region) and B phase (black region), respectively. 
To compute eq.~\ref{CH}, we used the Euler method and set a grid size to $\Delta x =1$ and a time increment to $\Delta t = 0.01$.
We used a free surface condition in the $x$ directions and a periodic boundary condition in the $y$ directions. 
The system size is $L_x : L_y = 1000 : 1000$.

We initially put A phase in $x < 40$ and then we put one or two slits of B phase in A phase as the initial perturbations (see Fig.~\ref{pattern}). 
The width of one slit or the period of two slits is $h$. 
The region $x \geq 40$ is a symmetric composition $\phi = 0$. 
Then, the quenching front was set to $x$ = 40 and annealed for $-10 < t < 0$ to smooth the slit boundary of $\phi$. 
Even after the annealing, the width of the slit was confirmed to be $h$. 
At $t$ = 0, the quenching front moved in the $x$ direction with a constant velocity $V$. 
Therefore, the temperature $\epsilon (x, t)$ is as follows. 
\begin{eqnarray}
\epsilon (x,t) = \left\{
\begin{array}{ll}
1 & x \geq Vt + 40\\
-1 & x < Vt +40
\end{array}
\right.
, \  \ \ \  x \in [0, L_x-1]
\end{eqnarray}

\section{Results}
\begin{figure}[htbp]
\centering
\includegraphics[width=8cm]{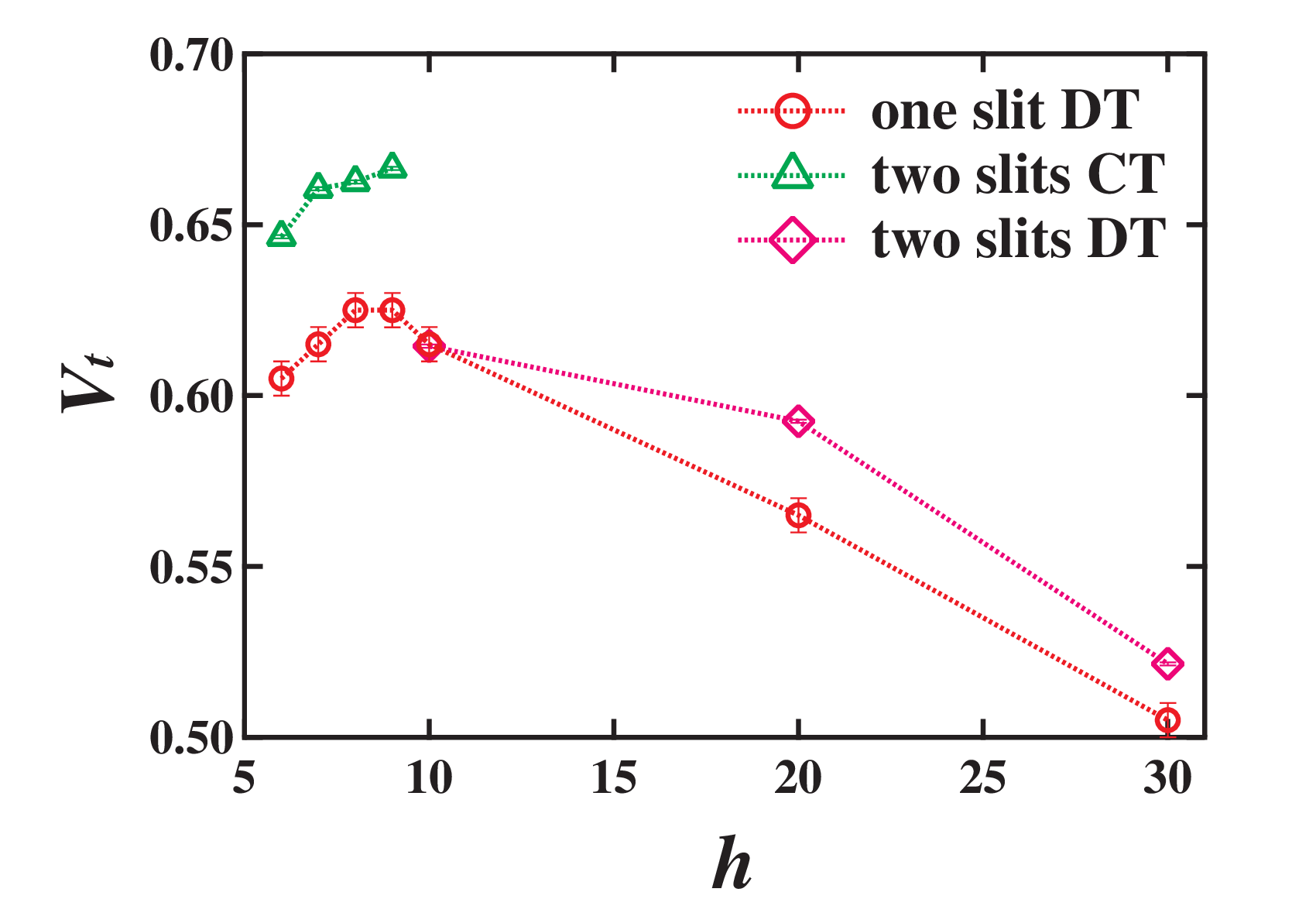}
\caption{$h$ dependence of $V_t$. Circles, triangles, and diamonds correspond to $V_t$ with a discontinuous transition (DT) for one slit, with a continuous transition (CT) for two slits, and with a DT for two slits, respectively. 
We set the transition velocity $V_t$ at the median of the lowest velocity for the lamellar pattern and the highest velocity for the columnar pattern. 
An error bar represents the lowest velocity for the lamellar pattern and the highest velocity for the columnar pattern. } 
\label{diagram}
\end{figure}

Firstly, we describe the pattern formation when the quenching front is moved at $V$ in the slit systems. 
We defined ``columnar pattern" when columns percolate to the right boundary of the simulation box, while ``lamellar pattern" is stable when the columns change into the lamellae during the pattern formation. 
We set the transition velocity $V_t$ at the median of the lowest velocity for the lamellar pattern and the highest velocity for the columnar pattern.
Figure~\ref{diagram} shows the $h$ dependence of $V_t$ in one or two slits system. 
There are two types of the transitions; one is a discontinuous transition (DT) and the other is a continuous transition (CT). 
We will describe the difference between DT and CT in the next paragraph. 
Circles, triangles, and diamonds correspond to $V_t$ with DT for one slit, with CT for two slits, and with DT for two slits, respectively. 
For the one slit system, the $V_t$ line is smooth with a peak at $h$ = 9 and the CT cannot be observed.  
Meanwhile, for the two slits system, $V_t$ is larger for $h \le 9$ (triangle symbols) than for one slit system.
For $h \ge 10$ (diamond symbols), $V_t$ with DT was almost the same as $V_t$ with DT in the one slit system. 
In the following, we will reveal the reason for the large $V_t$ jump around $h$ = 9 in the two slits case and we will discuss the difference from the one slit case. 

\begin{figure*}[htbp]
\centering
 \includegraphics[width=16cm]{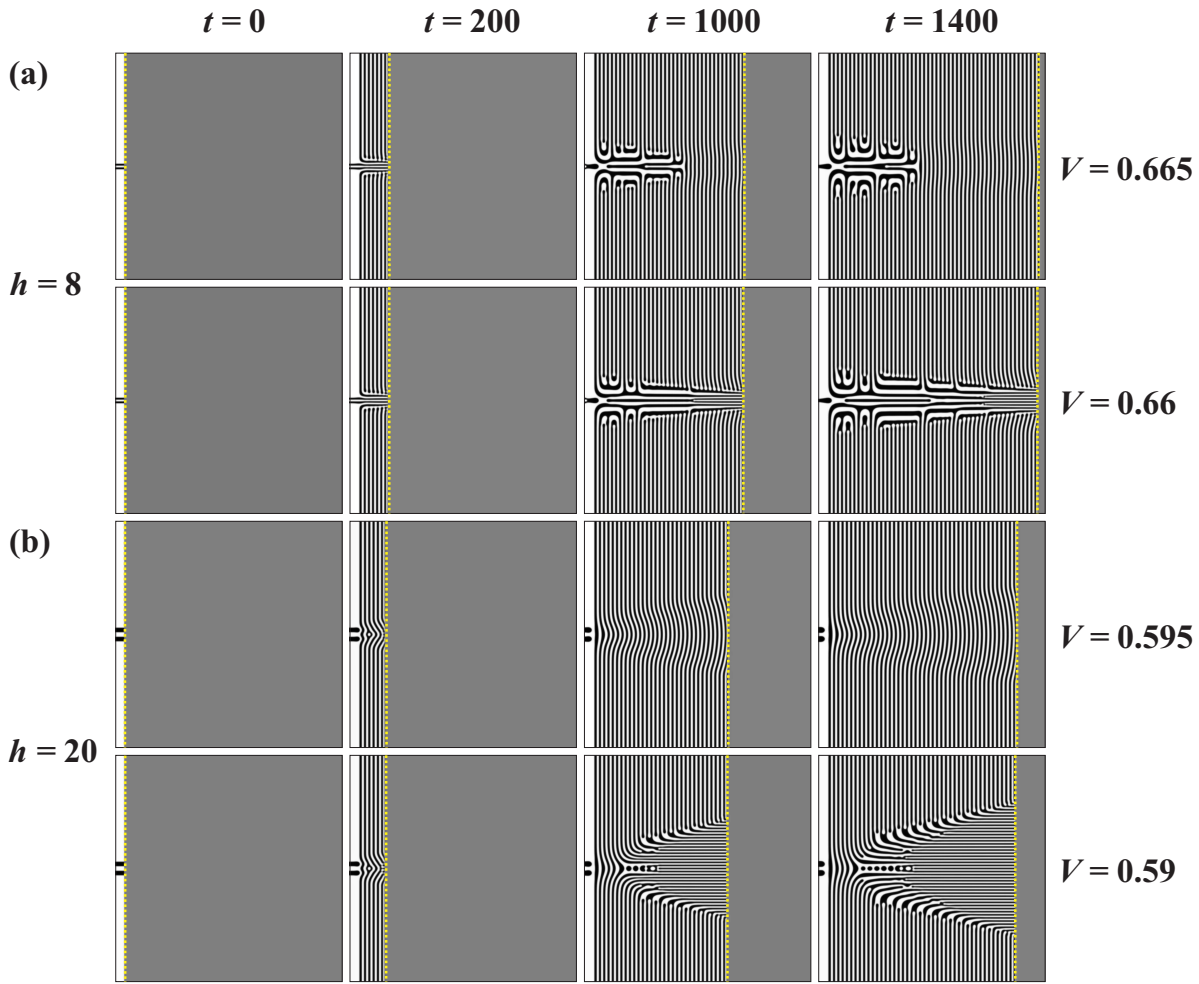}
  \caption{Time evolution of patterns in two slits. The yellow dotted lines represent the quenching front. 
(a) For $h$ = 8 and $V$ = 0.665 just above $V_t$, the number of the slit increases as columns, but they break off in the middle and lamellae are formed; the column formation has a finite lifetime (or a finite distance). 
Meanwhile, for $h$ = 8 and $V$ = 0.66 just below $V_t$, the columnar pattern percolate to the right boundary and the number of the column gradually increases. 
(b) For the slits width $h$ = 20 and $V$ = 0.595 just above $V_t$, no column is generated from the slits and only lamellae are formed immediately.  
Meanwhile, for $h$ = 20 and $V$ = 0.59 just below $V_t$, a thin column is formed at $t$ = 200 and then the number of the thin column increases and the columnar pattern percolates to the right boundary. 
}
  \label{pattern}
\end{figure*}

Next, we describe $h$ dependence of the pattern formation in the two slits case. 
Figure~\ref{pattern} shows the pattern formations for different $h$. 
For $h$ = 8 and $V$ = 0.665 just above $V_t$, the number of the slit increases as columns, but the columns break off in the middle and the lamellae are formed; the column formation have a finite lifetime (or a finite distance). 
Meanwhile, for $h$ = 8 and $V$ = 0.66 just below $V_t$, the columnar pattern percolates to the right boundary and the number of the column gradually increases. 
For the slits width $h$ = 20 and $V$ = 0.595 just above $V_t$, no column is generated from the slits and only lamellae are formed immediately after the DQ starts. 
Meanwhile, for $h$ = 20 and $V$ = 0.59 just below $V_t$, a thin column is formed at $t$ = 200 and then the number of the thin column increases and the columnar pattern percolates to the right boundary. 
Here we defined the continuous transition (CT) when the columns and lamellae coexist with some length like the system with $h$ = 8, while the discontinuous transition (DT) when the lamellae are formed immediately.

\begin{figure*}[htbp]
\centering
\includegraphics[width = 16cm]{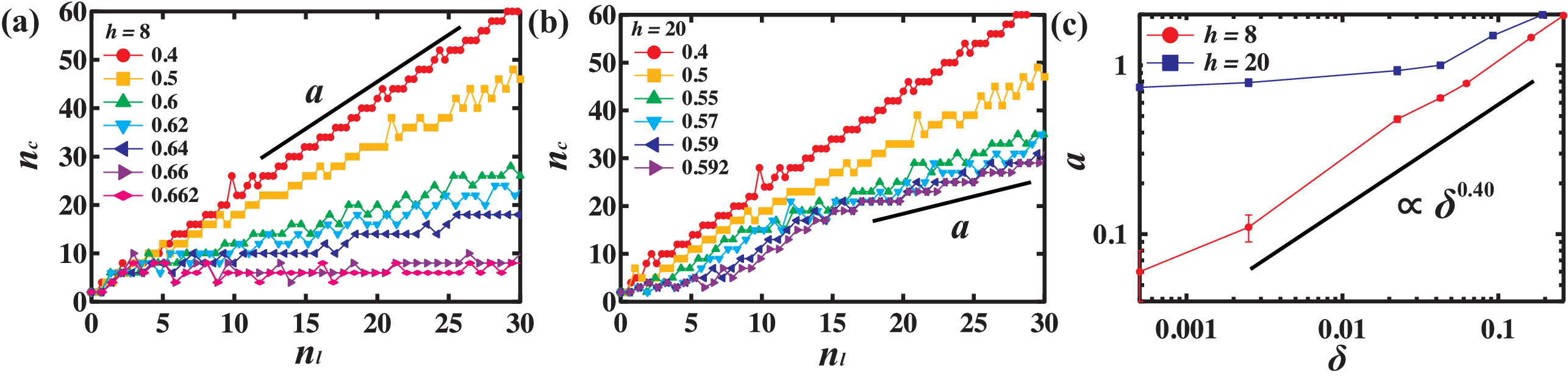}
\caption{$n_c$ dependence on $n_l$ for (a) $h$ = 8 and for (b) $h$ = 20.
Each symbols represent $n_c$ in a different $V$. 
For $h$ = 8, $n_c$ linearly increases at any $V$ and the slope decreases with increasing $V$.
For $h$ = 20, the slope change at the time of the changing from thick slits to thin column when $V$ is close to $V_t$ and the slope in the later period are almost constant near $V_t$. 
(c) $\delta (= V_t - V)$ dependence of the slope $a$ in log-log scales. Circles and squares correspond to $h$ = 8 and 20, respectively. 
For $h$ = 8, $a$ obeys with $\delta^\alpha$ and the power $\alpha$ is about 0.40. 
For $h$ = 20, $a$ has a finite value even $\delta \rightarrow 0$. 
The $a$ dependence of $\delta$ also suggests the difference between CT and DT modes.}
\label{a}
\end{figure*}

In addition, to clarify the mode of the transition, we investigated the growing behavior of the column pattern just below $V_t$. 
When the quenching front migrates toward $x$ direction, the lamellae are basically formed, but the column pattern also grows inside the lamellae (see Fig.~\ref{pattern}). 
We defined a lamella and a column as one pair of A and B phase, and the number of lamellae and columns as $n_l$, $n_c$, respectively. 
Figure~\ref{a} (a) and (b) show the $n_c$ dependence of $n_l$ for $h$ = 8 and 20, respectively. 
For $h$ = 8, $n_c$ linearly increases at any $V$. 
When $V$ is small, the slope $a$ is about 2, which indicates that two columns are formed each time one lamella formed. 
As $V$ approaches $V_t$, the slope decreases. 
Meanwhile, for $h$ = 20, the slope changes when the initial thick slit changes to the thin column near $V = V_t$. 
It is also found that the slopes in the later period are almost constant. 
We investigate the $a$ dependence on the velocity difference from $V_t$ $\delta (= V_t - V)$ shown as in Fig.~\ref{a}(c). 
Circles and squares correspond to $a$ in $h$ = 8 and 20, respectively. Error bars reflect the errors in the linear fitting. 
For $h$ = 8, $a$ obeys with $\delta^\alpha$ and the power $\alpha$ is about 0.40. 
The physical meaning of $\alpha$ is unclear, thus the theoretical derivation of this relationship is a future work.  
For $h$ = 20, $a$ has a non-zero value even $\delta \rightarrow 0$. 
Thus, the growth behaviors of the column pattern depend on $\delta$ and there is a clear difference between CT ($h$ = 8) and DT ($h$ = 20). 

\begin{figure}[htbp]
\centering
\includegraphics[width=8cm]{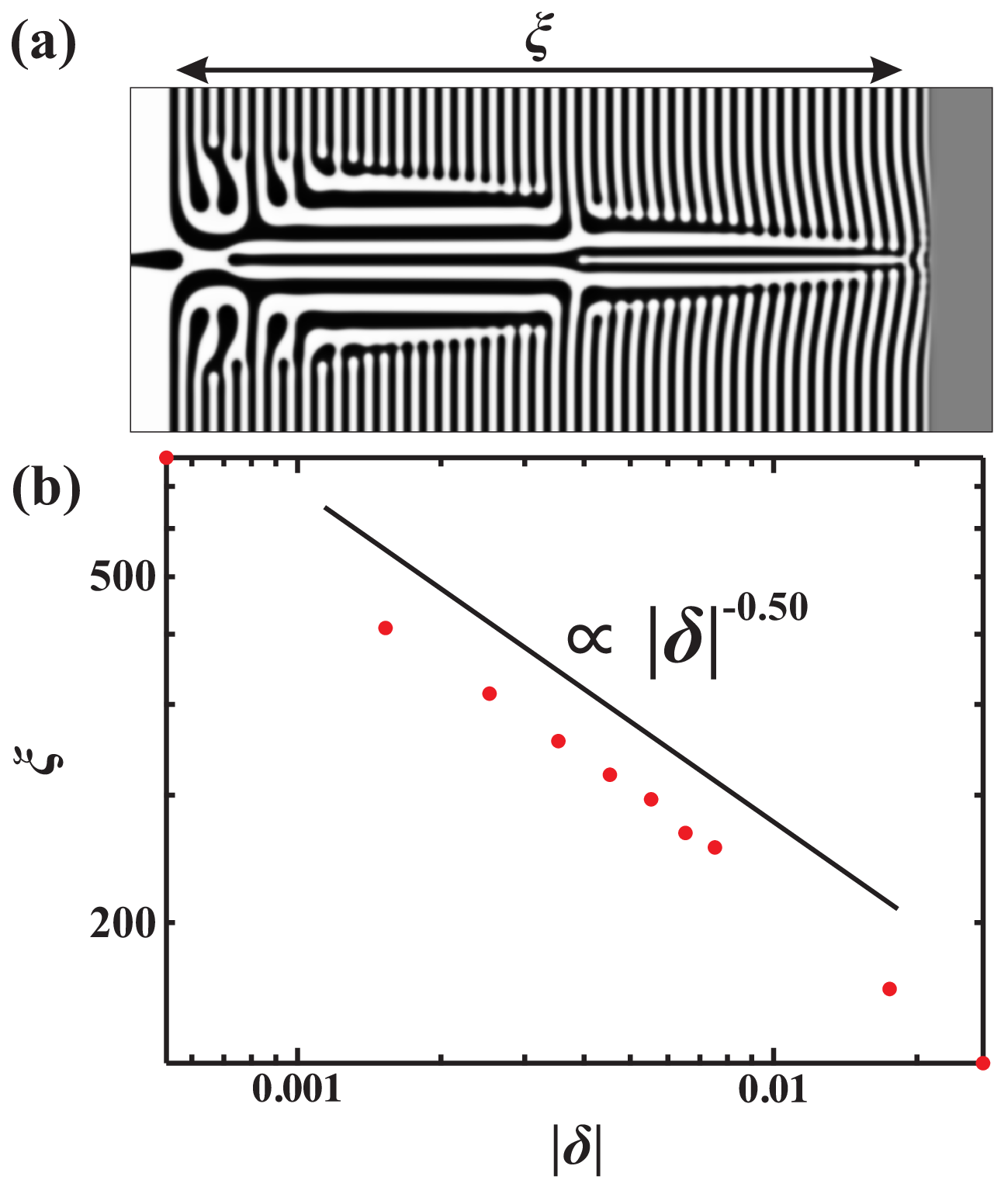}
\caption{(a) We define $\xi$ as the distance from the initial position of the slit to the vanishing point of the columns at $h$ = 8. (b) $|\delta|$ dependence of $\xi$. $\xi$ diverges as $\xi \sim |\delta|^{\beta}$ with decreasing $|\delta|$ and $\beta$ is about -0.50. This behavior also suggests that the pattern transition at $V = V_t$ should be continuous at $h$ = 8. }
\label{lifetime}
\end{figure}

We also show the pattern formation behaviors above $V_t$. 
To clarify the dynamics in CT, we investigated the coexistence length of the columns and the lamellae for $h$ = 8. 
We define $\xi$ as the distance from the initial position of the slit to the vanishing point of the columns (see Fig.~\ref{lifetime} (a)) and investigate $\xi$ dependence on $|\delta| (=V - V_t)$ shown as in Figure~\ref{lifetime}(b).
$\xi$ diverges as $\xi \sim |\delta|^{\beta}$ with decreasing $|\delta|$ and $\beta$ is about -0.50. 
This behavior also suggests that the pattern transition at $V = V_t$ should be continuous at $h$ = 8. 
Meanwhile, $\xi$ = 0 at $h$ = 20 as we noted above (see Fig.~\ref{pattern}(b). 

\begin{figure}[htbp]
\centering
\includegraphics[width=8cm]{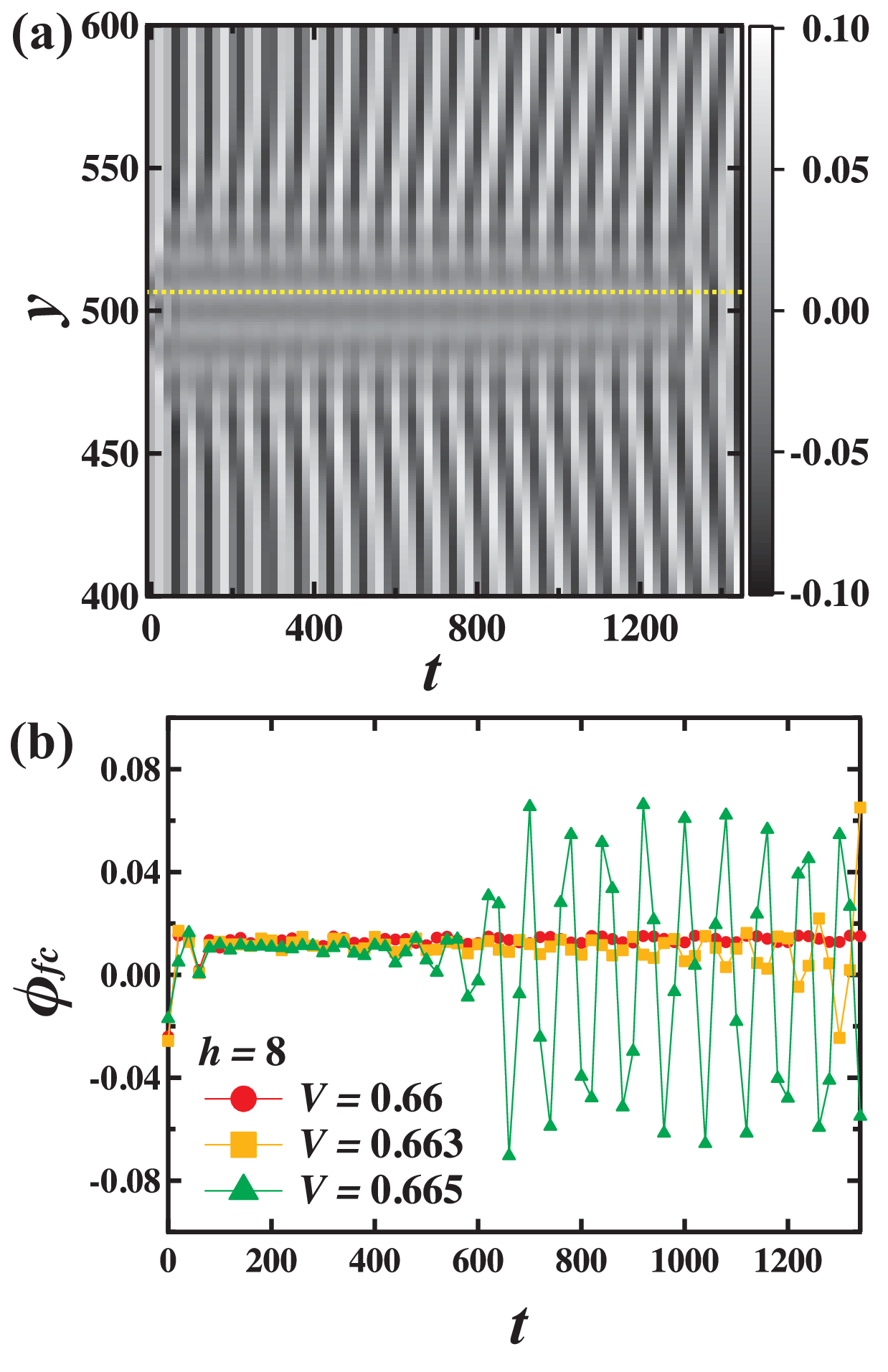}
\caption{(a) The time evolution of $\phi (x_f+3, y)$ at $h$ =8 and $V$ = 0.663, where $x_f$ is a position of the quenching front at $t$. 
 $x_{f}$ and $y_{c}$ show the $x$-coordinate of the quenching front and the $y$-coordinate of the column center (indicated by yellow dotted line), respectively. 
(b) Time evolution of $\phi_{fc}$ at $h$ = 8. Circles, squares and triangles are $V$ = 0.66, 0.663, and 0.665, respectively. 
$V_t$ = 0.6625 at $h$ = 8. 
When $V$ is smaller than $V_t$ = 0.6625 ($V$ = 0.66), the amplitude of $\phi_{fc}$ oscillation is quite small at any $t$. 
Meanwhile, when $V$ is larger than $V_t$ ($V$ = 0.663 and 0.665), the oscillations of $\phi_{fc}$ become larger with time and then the column finally transforms into lamellae. The dynamics in CT is determined by the competition between the accumulation and the diffusion in front of the column. }
\label{profile}
\end{figure}

Here we investigated the time evolution of the concentration at a position prior to the quenching front of the column in order to reveal the formation dynamics in the continuous transition. 
Figure \ref{profile}(a) shows the time evolution of $\phi (x_f+3, y)$ at $h$ =8 and $V$ = 0.663, where $x_f$ is a position of the quenching front at $t$. 
We also define the $y$-coordinate of the center of the column as $y_c$ (indicated by yellow dotted line) and $\phi (x_f+3, y_c)$ is denoted by $\phi_{fc}$ in the following.
Since the concentration is conserved and A phase rich ahead of the column with B phase, $\phi_{fc}$ is positive (A rich phase), while the column is composed by B phase. 
As time passed, $\phi_{fc}$ periodically changes shown as in Fig.~\ref{profile}(b). 
Circles, squares and triangles are $V$ = 0.66, 0.663, and 0.665, respectively. 
When $V$ is smaller than $V_t$ = 0.6625 ($V$ = 0.66), the amplitude of $\phi_{fc}$ oscillation is quite small at any $t$. 
Meanwhile, when $V$ is larger than $V_t$ ($V$ = 0.663 and 0.665), the oscillations of $\phi_{fc}$ become larger with time and then the column finally transforms into lamellae. 

Here, we considered the periodic fluctuations and divergence of $\phi_{fc}$.  
Due to the conserved manner, A component accumulates in front of the B phase columns ($\phi_{fc} > 0$ in front of the B column). 
When B phase columns grow, the accumulated A component is diffused in the lateral direction. 
When a lamella of A phase is formed next to B phase column, a layer with $\phi < 0$ is formed in front of the quenching front. 
Then the accumulated A component in front of the B phase columns tends to diffuse the laterally since the difference of $\phi$ is large.  
Thus, $\phi_{fc}$ oscillates slightly with the same period as the period of the lamella. 
Next, the quenching front movement is faster than the characteristic time of the diffusion when $V \ge V_t$. 
The A component is gradually accumulated in front of the B phase columns with time. 
The amplitude of the oscillation gradually increases and the onset time, when the amplitude of $\phi_{fc}$ oscillation becomes larger, becomes shorter with increasing $V$.   
They are consistent that the accumulation of A component is faster than the diffusion. 
Therefore, the dynamics in CT is determined by the competition between the accumulation and the diffusion in front of the column. 

\begin{figure}[htbp]
\centering
\includegraphics[width=8cm]{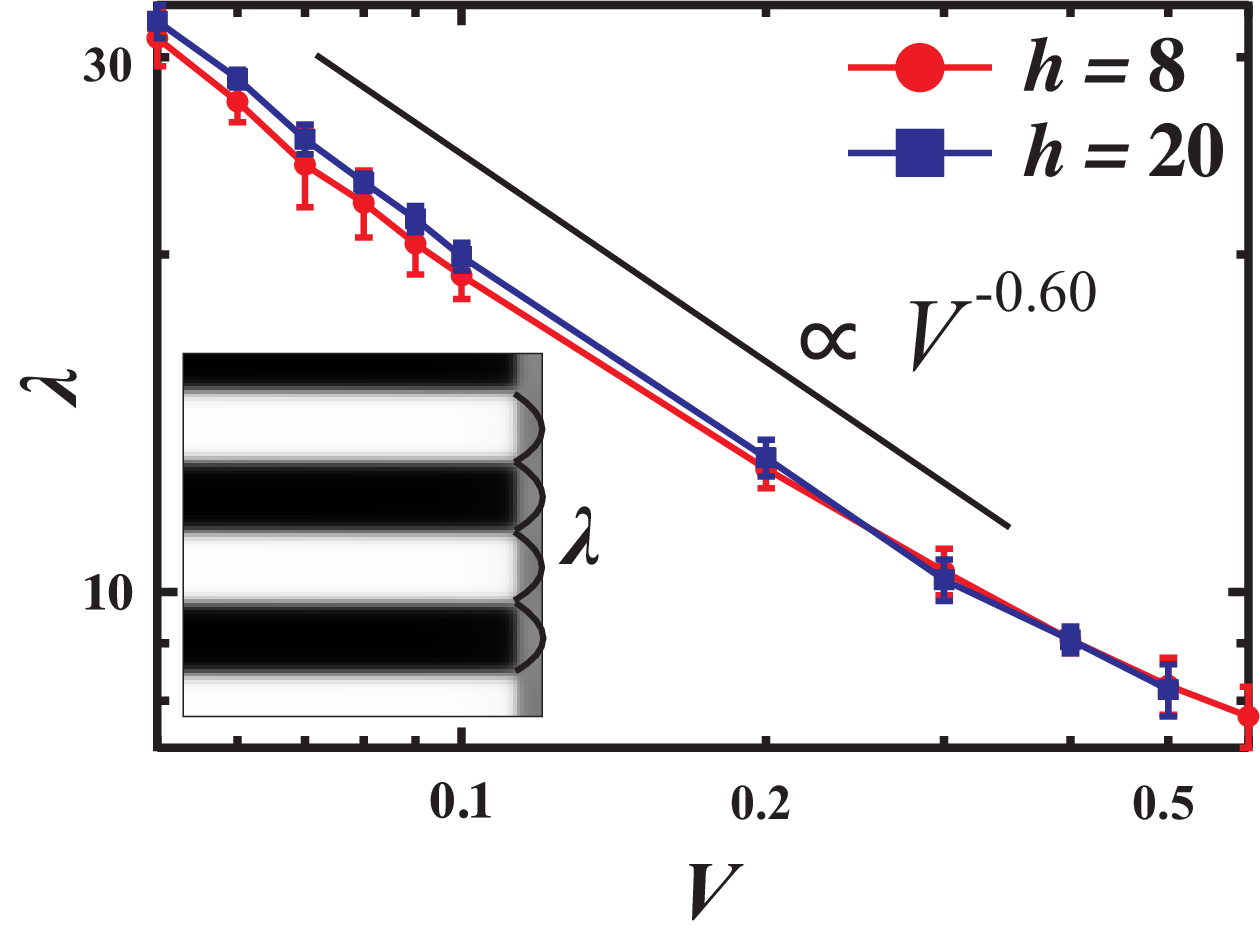}
\caption{$V$ dependence of $\lambda$ just after the entire system is quenched. Circles and triangles are $h$ = 8 and 20, respectively. Symbols and error bars are the mean and standard deviation of $\lambda$ for all columns, respectively. It is found that $\lambda$ is independent of $h$ and $\lambda \propto V^{-0.60}$. }
\label{lambda}
\end{figure}

When the width of the initial slit $h$ is large, the thinner column is reformed during DQ.  
It means that the stable width of the column is mismatched with the width of the initial slit.
Thus, we compute the stable width of the column here. 
According to the previous report in Ref.~\cite{Ishikawa2023}, a stable thickness of the lamella $\xi^{\prime}$ can be expressed as $\xi^{\prime} \sim V^{\prime -1/2}$ since the time scales of the diffusion and the migration of the free surface are balanced. 
Similarly, the stable width of the column is expected to be determined by the competition between the velocity of the quenching front and the speed of the diffusion. 
We investigated the $V$ dependence of the stable width of the column $\lambda$ just after the entire system was quenched. 
Figure \ref{lambda} shows the $V$ dependence of $\lambda$. Circles and triangles are $h$ = 8 and 20, respectively. 
Symbols and error bars are the mean and the standard deviation of $\lambda$ for all columns, respectively. 
It is found that $\lambda$ is independent of $h$ and $\lambda \propto V^{-0.60}$. 
In the lamella pattern, the diffusion perpendicular to the quenching front is only considered, while the diffusion in both perpendicular and lateral direction should be related in the column pattern. 
Thus we consider that the exponent in the column pattern is different from that in the lamella pattern. 
When $h$ is large and $V < V_t$, the wide initial slit branches into columns of the steady width $\lambda$. 
Meanwhile, when $V \ge V_t$, the diffusion is slower than the migration of quenching front. 
A layer of A component is formed in front of the B phase column and then the layer becomes the stable lamellar pattern. 

\begin{figure}[htbp]
\centering
\includegraphics[width=8cm]{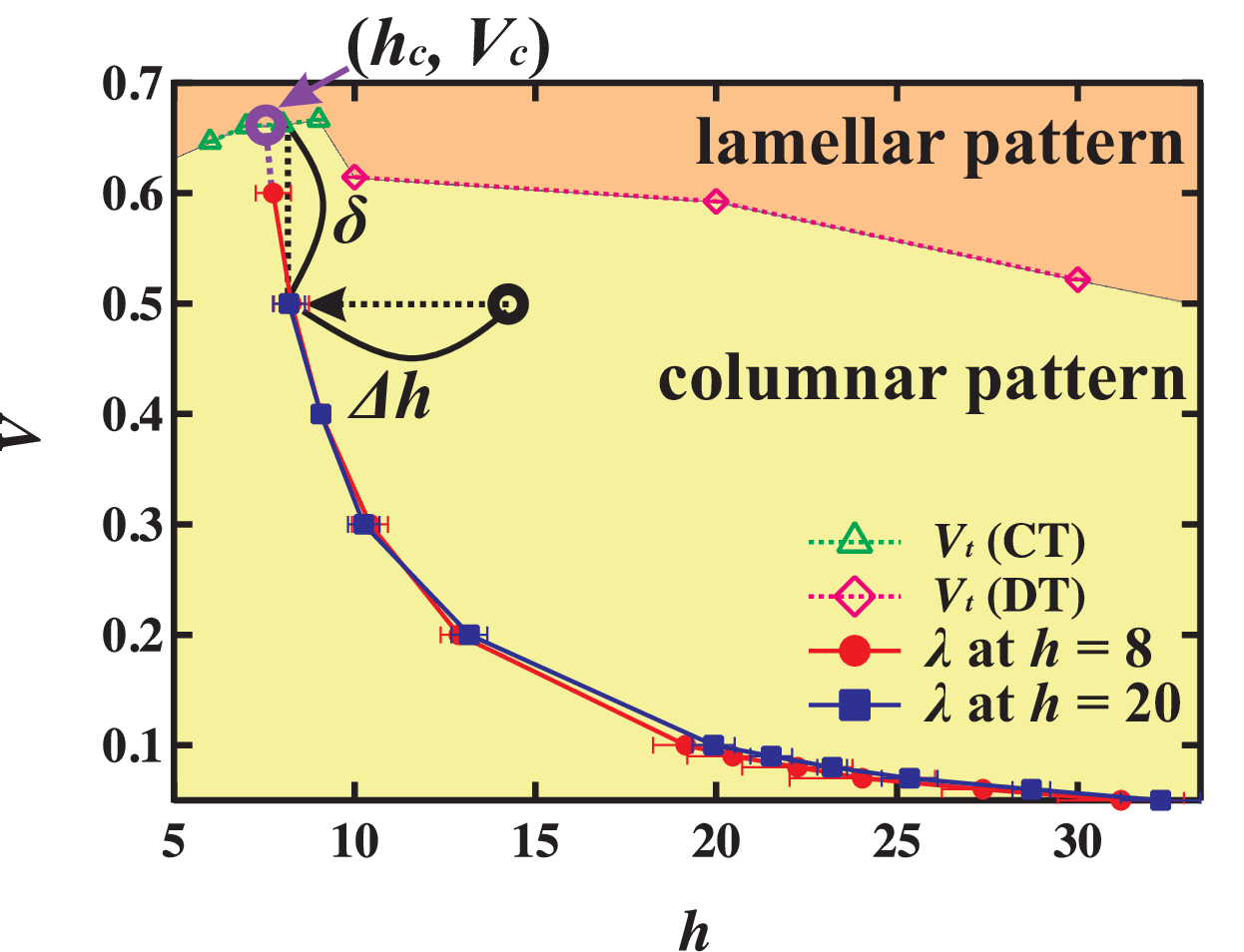}
\caption{A superimposed graph of Fig.~\ref{diagram} and Fig.~\ref{lambda} with $V$ and $h$. The vertical and horizontal axis of Fig.~\ref{lambda} are swapped. 
Filled circles and filled squares are the column width $\lambda$ at $h$ = 8 and 20, respectively. 
There exists a critical-like point $(h, V) = (h_c, V_c)$, where $\lambda$ and $V_t$ curves cross. 
When the condition is close to the critical-like point, the transition is continuous. 
In the case $h$ = 15 and $V$ = 0.5 (black circle), a misfit parameter between the initial slit and the stable column $\Delta h = h - \lambda$ is large.
When $V < V_t$, the slits for $h$ = 15 branch to a column with $\lambda = 8$ to eliminate this misfit.  
In the post-bifurcation state, $\delta = V_t - V$ is also large and then the transition is discontinuous. 
}
\label{discuss}
\end{figure}

Finally, we explain the conditions under which CT occurs. 
Figure \ref{discuss} is a superimposed graph of Fig.~\ref{diagram} and Fig.~\ref{lambda} with $V$ on the vertical axis and $h$ on the horizontal axis. 
Filled circles and filled squares are the stable column width $\lambda$ at the initial slit width $h$ = 8 and 20, respectively. 
There is a critical-like point $(h, V) = (h_c, V_c)$, where $\lambda (V_c) = h_c$ and $V_t (h_c) = V_c$. 
Figure~\ref{discuss} is a state diagram and the transition modes between CT and DT can be explained by this diagram. 
As an example, we consider the case $h$ = 15 and $V$ = 0.5 (black circle). 
In this case, a misfit parameter between the initial slit and the stable column $\Delta h = h - \lambda$ is large.
When $V < V_t$, the slits for $h$ = 15 branch to a column with $\lambda = 8$ to eliminate this misfit.  
In the post-bifurcation state, $\delta = V_t - V$ is also large and the ratio of column formation to lamella $a$ is also large. 
Meanwhile, the slits immediately become a lamella if $V > V_t$. 
Therefore, due to the bifurcation, the condition cannot be close to the critical-like point and then the transition is discontinuous. 
Meanwhile, $\lambda \sim h_c$ holds at $h$ = 8 and $V = V_c$. 
In this case, the condition of DQ is close to the critical-like point and the transition becomes continuous. 
We note here that if $h$ is slightly larger than $\lambda$, the slits cannot branch. Therefore, we consider that the transition is the continuous up to $h$ = 9. 

In addition, we describe the difference of the transition mode between one slit and two slits. 
One slit is more likely to be lamellae than two slits because the upper and lower lamellae are closer together. 
Therefore, $V_t$ for one slit is smaller than for two slits. 
If $V < V_t$, two or more columns are formed from one slit. 
When two or more columns are formed, the stability increases and then $V_t$ becomes larger. 
Then, the condition becomes far from the critical-like point. 
We note that the state diagram for three or more slits is similar to that of two slits. 
Therefore, the stability of the pattern is unchanged after two or more columns are formed.

These results suggest that the stability of the system changes with the time dependent condition such as the number of the slit, the width of the slit or the column. 
Then the discontinuous transition occurs since the re-established transition point is farther away from the simulated state.
Conversely, if the initial inhomogeneous perturbations are close to the stable state, the stability of the system is unchanged and then the continuous transition occurs near the critical point. 
It can be concluded that the difference between the initial inhomogeneous perturbations (the slit width $h$ in this study) and the steady state (the column width $\lambda$ in this study) significantly affects the mode of the pattern formation. 

\section{Discussion}
Firstly, we discuss the relevance of this study with experimental studies on directional pattern formation during eutectic growth and gelation process. 
In metallic alloys, eutectic patterns have been observed to self-assemble perpendicular to the crystal-growth direction (like columnar pattern) when it solidifies in a certain direction~\cite{Ginibre1997, Jackson-hunt1966, Kulkarni2018}. 
However, a horizontal pattern (like lamellar pattern) is not formed with respect to the crystal-growth direction. 
Similarly, in gelation of collagen induced by directional neutralization, a pattern perpendicular to the gelation plane (like columnar pattern) is formed \cite{Yonemoto2021}, but a horizontal pattern (like lamellar pattern) is not formed. 
In this study, $V_t$ with the initial perturbations is about one order of magnitude larger than $V_t$ in homogeneous DQ \cite{Tsukada2020} and then the $V$ regime of the lamellar pattern becomes narrow. 
Therefore, in nonequilibrium systems, the uncontrollable initial perturbations exist and then it seems difficult to observe lamellar patterns in experiments. 

Next, we discuss the relationship between CT in DQ and the general critical phenomena. 
The results of the continuous transition in DQ are deterministic, not taking into account thermal fluctuations. 
On the other hand, the critical phenomena of the phase transition~\cite{Onuki2002} and a laminar turbulent transition~\cite{Sano2016} are stochastic with thermal fluctuations. 
In fact, the columns do not disappear at the center of the columnar pattern, which is different from a critical phenomenon. 
Therefore, in this stage, the CT in DQ is different from these critical phenomena, and it is more appropriate to consider the pattern transition as interfacial dynamics. 

Finally, the transition of the pattern formation under inhomogeneous conditions are not limited to phase separations, but also occur in other nonequilibrium systems such as active matter~\cite{Vicsek1995, Ginelli2016, Chate2020}, non-reciprocal phase transitions~\cite{Fruchart2021}, reversible-irreversible transitions in jamming systems~\cite{Nagasawa2019, Reichhardt2023}. 
It is interesting that nonequilibrium continuous transition are observed in many systems such as the laminar-turbulent transition in fluids~\cite{Sano2016}, two turbulent states in liquid crystals~\cite{Takeuchi2007, Takeuchi2009} and a reversible-irreversible transition of fiber~\cite{Franceschini2011}, vortices~\cite{Maegochi2021, Mangan2008} and granular materials~\cite{Ness2020}. 
Thus it is expected that the difference between the condition at the steady state and the initial perturbations plays an important role in those nonequilibrium transition. 

\section{Summary}
A phase separation is important in a wide range of fields such as material science and biology. 
Those phase separations often exhibit in inhomogeneous concentrations or non-stationary temperature fields. 
Understanding of phase separation phenomena under such inhomogeneous conditions is an urgent issue since LLPS in cells occurs in inhomogeneous concentration and temperature and the pattern formation are related with development of new functional materials.

Here, we investigated the columnar-lamellar pattern transition due to directional quenching (DQ) in the presence of initial slits pattern to study the effect of initial perturbations. 
In this study, we prepared one slit and two slits which are the nuclei of the columnar pattern as initial perturbations. 
Then, we studied the process of transition from columnar pattern to lamellar pattern dependent on the initial perturbations. 

It is found that the transition behavior depends on the slit width $h$ and the number of the slits. 
The transition from the column to the lamella is continuous when $h$ is comparable to the column width $\lambda$ and two slits. 
Then, the growth of the columnar pattern is determined by the balance between the diffusion and the accumulation of the concentration. 
Whereas, it is a discontinuous transition when $h$ is larger than the column width $\lambda$. 
The stability of the system changes with the time dependent condition such as the number of the slit, the width of the slit or the column. 
Then the discontinuous transition occurs since the re-established transition point is farther away from the simulated state.

This study not only shows the empirically knowledge that inhomogeneous initial perturbations can significantly change the transition point from a homogeneous system. 
It also shows that in nonequilibrium systems, the transition behaviors are determined by the difference between the initial perturbations and the condition at the steady state. 
Therefore, it is important to design initial perturbations comparing with the condition at the steady state in order to control pattern formation and nonequilibrium transitions in experiments.

\section*{Acknowledgements}
R. I. was supported by JST SPRING, Grant Number JPMJSP2156. R. K. was supported by JSPS KAKENHI Grant Number 20H01874. 

\section*{AUTHORS CONTRIBUTIONS}
R.~I.  and R.~K. conceived the project. R.~I. performed the numerical simulations and analyzed the data. R.~I. and R.~K. wrote the manuscript.

\section*{COMPETING INTERESTS STATEMENT}
The authors declare that they have no competing interests. 

\section*{CORRESPONDENCE}
Correspondence and requests for materials should be addressed to R.~I. (ishikawa-rikuya@tmu.ac.jp) and R.~K. (kurita@tmu.ac.jp).

\section*{Availability of Data and Materials}
All data generated or analyzed during this study are included in this published article and its supplementary information files.

\end{document}